\documentclass[twocolumn,prl,aps,amssymb,amsmath,superscriptaddress,showpacs,
showkeys,floatfix,a4paper]{revtex4}
\usepackage{graphicx}

\begin{document}
\title{Effects of atomic interactions on Quantum Accelerator Modes.}

\author{Laura Rebuzzini}
\email{laura.rebuzzini@uninsubria.it}
\affiliation{Center for Nonlinear and Complex Systems and
Dipartimento di Fisica e Matematica, Universit\`a dell'Insubria, 
Via Valleggio 11, 22100 Como, Italy.}
\affiliation{Istituto Nazionale di Fisica Nucleare, Sezione di Pavia, 
Via Ugo Bassi 6, 27100 Pavia, Italy.}
\author{Roberto Artuso}
\affiliation{Center for Nonlinear and Complex Systems and
Dipartimento di Fisica e Matematica, Universit\`a
dell'Insubria, Via Valleggio 11, 22100 Como, Italy.}
\affiliation{Istituto Nazionale di Fisica della Materia, 
Unit\`a di Como, Via Valleggio 11, 22100 Como, Italy.}
\affiliation{Istituto Nazionale di Fisica Nucleare, Sezione di Milano,
Via Celoria 16, 20133 Milano, Italy.}
\author{Shmuel Fishman}
\affiliation{Physics Department,  Technion, Haifa 32000, Israel.}
\author{Italo Guarneri}
\affiliation{Center for Nonlinear and Complex Systems and
Dipartimento di Fisica e Matematica, Universit\`a
dell'Insubria, Via Valleggio 11, 22100 Como, Italy.}
\affiliation{Istituto Nazionale di Fisica Nucleare, 
Sezione di Pavia, Via Ugo Bassi 6, 
27100 Pavia, Italy.}
\affiliation{Istituto Nazionale di Fisica della Materia, 
Unit\`a di Como, Via Valleggio 11, 22100 Como, Italy.}


%
\begin{abstract}
{We consider the influence of the inclusion of interatomic interactions 
on the $\delta$-kicked accelerator model. 
Our analysis concerns in particular quantum accelerator modes, namely quantum ballistic transport near quantal resonances. 
The atomic interaction is modelled by a Gross-Pitaevskii cubic nonlinearity, 
and we address both attractive (focusing) and repulsive (defocusing) cases. 
The most remarkable effect is enhancement or damping of the accelerator modes, 
depending on the sign of the nonlinear parameter.  We provide arguments showing that the effect persists beyond 
mean-field description, and lies within the experimentally accessible 
parameter range.
}
\end{abstract}
\maketitle
\narrowtext

Quantum Accelerator Modes (QAMs) are a manifestation
of a novel type of quantum ballistic transport (in momentum), that has been recently observed in cold atom optics
\cite{Ox99}.
In these experiments, ensembles of about $10^7$ cold alkali 
atoms are cooled in a magnetic-optical trap 
to a temperature of a few microkelvin. 
After releasing the cloud, the atoms 
are subjected to the joint action of the gravity acceleration 
and a pulsed potential periodic in space, generated by 
a standing electromagnetic wave, far-detuned 
from any atomic transitions. The external optical potential is  
switched on periodically in time and the period is much longer 
than the duration of each pulse.
For values of the pulse period near to a resonant
integer multiple of half of a characteristic time $T_B$
(the Talbot time \cite{BB}), typical of the kind of atoms used, 
a considerable fraction of the atoms undergo a constant acceleration with 
respect to the main cloud, which falls freely under gravity and spreads
diffusively.

\begin{figure}
  \includegraphics[width=9cm,angle=0]{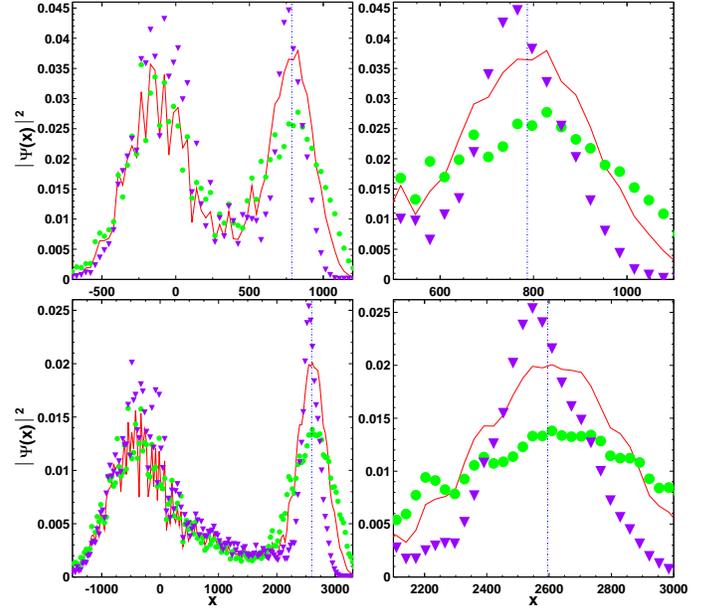}
  \caption{(Color online) The probability 
distribution at times $t=25$ ($1^{\rm st}$ row) and 45 ($2^{\rm nd}$ row). 
((Red) line: linear case ($u=0$),
(purple) triangles/(green) circles: focusing/defocusing nonlinearity
($u=\mp 1.25$)). In the right column enlargements of mode are shown;
the position of the mode, predicted by (\ref{acc}) 
is marked by the (blue) vertical dotted line.
}\label{pac1}
\end{figure}

The non-interacting model is a variant of the well-known quantum 
kicked rotor (KR) \cite{kr}, 
in which the effects of a static force, produced by the earth gravitational field, 
are taken into account.
The linear potential term breaks invariance of the KR hamiltonian under space translations. 
Such an invariance may be recovered by moving to a temporal gauge, where momentum is measured {\em w.r.t.} 
the free fall: this transformation gets rid of the linear term and the new hamiltonian, 
expressed in dimensionless units, reads
\begin{equation}
\label{ham-ff}
\hat H (t')=\frac 12 (\hat p+gt')^2+k\cos (\hat x)
\sum_{t=-\infty}^{t=+\infty} \delta (t'-t\tau ).
\end{equation}
where $\hat p$ and $\hat x$ are the momentum and position operator, 
$k$ and $\tau$ are the strength and the temporal period of the
external kicking potential, $g$ is the gravity acceleration.
The relationship between the rescaled parameters and the physical ones, 
denoted by primes, is $k=k'/\hbar$, $\tau=\hbar \tau ' G^2/M=4\pi \tau '/T_B$, 
$\eta=Mg'\tau '/\hbar G$ and $g=\eta/\tau$, where $\eta$ is the momentum gain over one period, 
$G$ is twice the angular wavenumber of the standing wave of the driving potential and $M$ is the mass of the atom.

Symmetry recovery allows to decompose the wavepacket into a bundle of independent rotors (whose space coordinate is topologically an angle): this Bloch-Wannier fibration plays an important role in the theory of QAMs \cite{FGR02}.

QAMs appear when the time gap between kicks approaches a principal quantum resonance, {\it i.e.} $\tau = 2 \pi l + \epsilon$, with $l$ integer and $|\epsilon|$ small. The key theoretical step is that in this case the quantum propagator may be viewed as the quantization of a classical map, with $|\epsilon|$ playing the role of an effective Planck's constant \cite{FGR02}: QAMs are in correspondence with stable periodic orbits of such pseudo-classical area-preserving map. We refer the reader to the original papers for a full account of the theory, we just mention a few remarkable points: stable periodic orbits are labelled by their action winding number $w=j/q$, which determines the acceleration of the QAM {\it w.r.t.} the center of mass distribution
\begin{equation}
\label{acc}
a\,=\,\frac{2 \pi}{|\epsilon|} \frac{j}{q} - \frac{\tau \eta}{\epsilon}.
\end{equation}
The modes are sensitive to the quasimomentum (Bloch index induced by spatial periodicity), being enhanced at specific, predictable values \cite{FGR02}; also the size of the elliptic island 
around the pseudoclassical stable orbit plays an important role 
(if the size is small 
compared to $|\epsilon|$ the mode is not significant \cite{FGR02}).

\begin{figure}
  \includegraphics[width=9cm,angle=0]{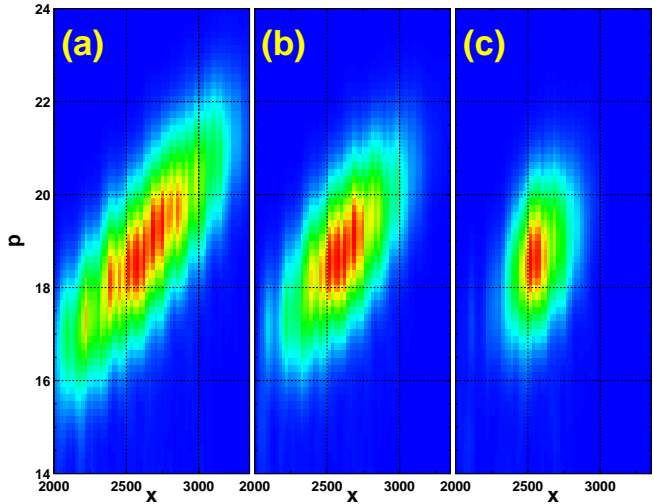}
  \caption{(Color online) The Husimi function of the QAM at 
time
$t=45$, in the repulsive (a), 
linear (b) and attractive case (c).}
\label{hus-z}
\end{figure}
\begin{figure}
  \includegraphics[width=9cm,angle=0]{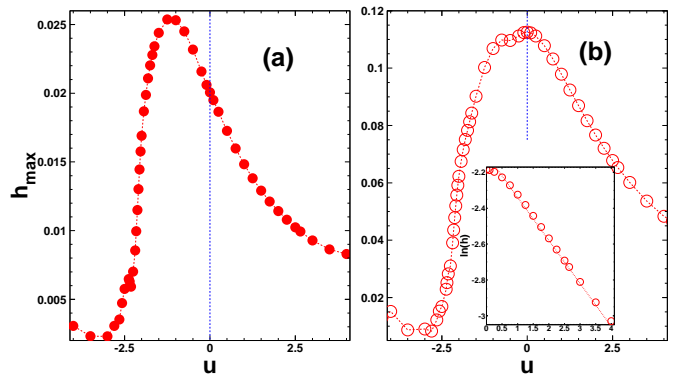}
  \caption{(Color online) Maximum height reached by the mode at time $t=45$
as a function of $u$ in position
(a) and momentum (b) representation.
 In the inset the exponential decrease of $h_{max}$ for positive $u$
in shown in a semi-logarithmic plot.}\label{hpm}
\end{figure}

We consider in this letter the role of atomic interactions in such a system; 
namely evolution is determined by a nonlinear Schr{\"o}dinger equation with a cubic nonlinearity:
\begin{equation}
\label{gp}
i{\dot\psi}(x,t')=\left[ 
\hat H (t')+u|\psi(x,t')|^2
\right]\psi(x,t'),
\end{equation}
where $u$ is the rescaled nonlinear 
parameter, whose sign describes an attractive (negative)/repulsive (positive) atomic interaction. 
We will come back to its connection with physical units in the end of the paper. 
 The condensate wave function is normalized to unity.
The dynamics does not only acquire in this way a qualitative novel form, but,  
 due to the nonlinear term, Bloch decomposition into independent rotors breaks down. 
The main scope of 
this letter will be to numerically scrutinize how QAMs are still present in the modified system, and explore how nonlinearity modifies their features. In the end we will briefly comment upon some stability issues, by showing that a more refined description, including loss of thermalized particles, does not destroy the scenario we get from a mean field description.

Our analysis will be restricted to QAMs corresponding to fixed points of period $q=1$ of 
the pseudoclassical map; the numerical analysis of nonlinear evolution has been 
performed by using standard time-splitting spectral methods \cite{tssm}.
There are several physical parameters characterizing the system: $g$, $\tau$, $k$ and $u$. Here we mainly address the role of nonlinearity $u$: 
we fix $k=1.4$, $l=1$, $\epsilon=-1$, $\tau \eta \simeq 0.4173$, 
and choose as the initial state a symmetric coherent state centered in the stable fixed point of 
the pseudoclassical map ($x_0\simeq 0.3027$, $p_0=0$), whose corresponding winding number is zero.

A quite remarkable feature appears when we compare results for opposite nonlinearity signs 
(keeping the strength $|u|$ fixed), see fig.(\ref{pac1}).
As in the linear system, the wave packet splits into two well-separated components: the accelerator mode (whose acceleration is still compatible with (\ref{acc})) and 
the remaining part, which moves under two competitive contributions, 
the free fall in the gravitation field, 
and the recoil against the accelerating part. Note that 
for the present choice of the parameters, the former contribution is 
negligible compared to the second.
\begin{figure}
  \includegraphics[width=9cm,angle=0]{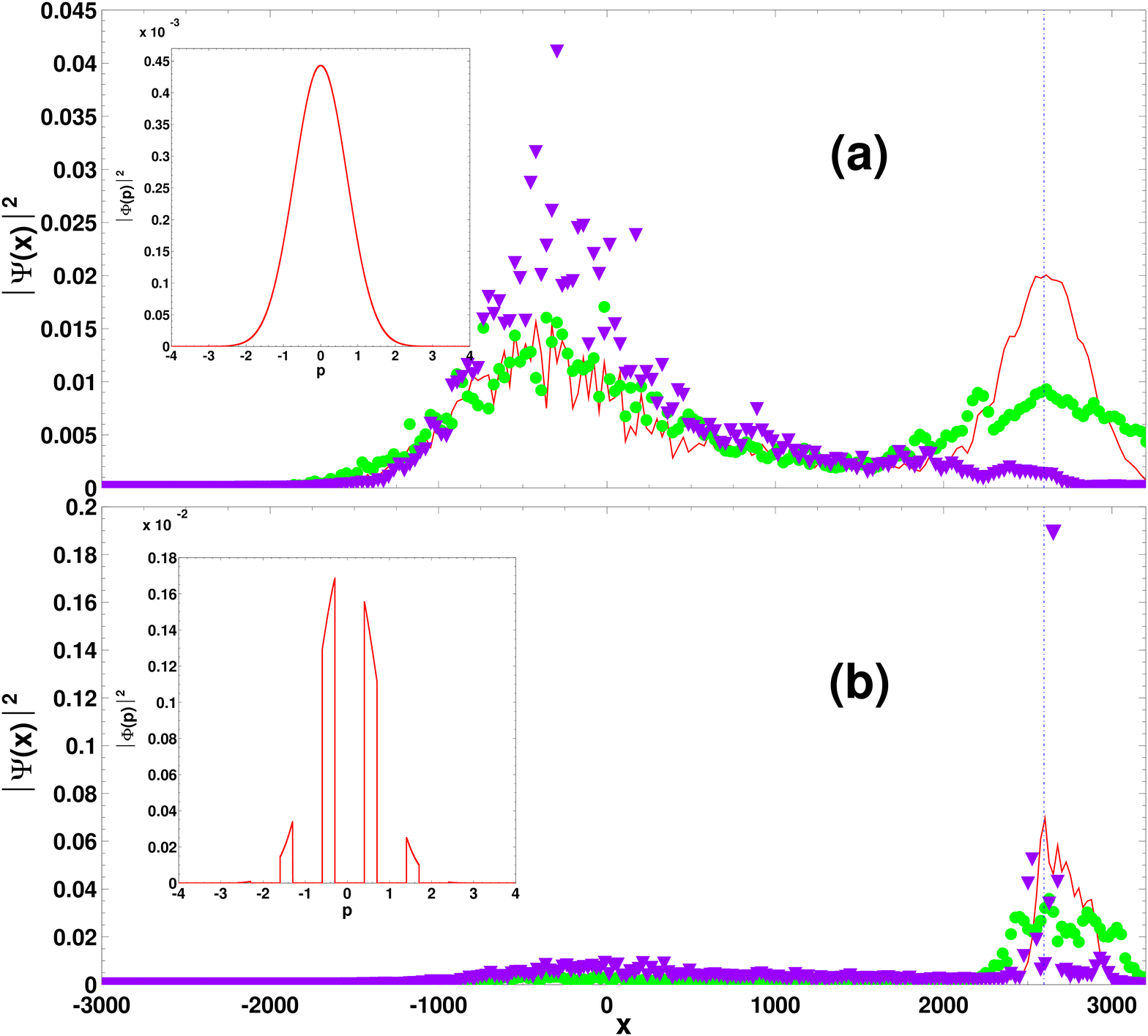}
  \caption{(Color online) The probability distribution at $t=45$ for 
strong nonlinearity ($u=\pm 3$); 
the initial states (see text) are shown in the insets 
(line and symbols as in fig.\ref{pac1}).}
\label{pac-x-ugr}
\end{figure}

We remark some features, that are common to what we observed for a 
choice of other parameter values: 
the distribution around the accelerator mode is more peaked and narrower 
in the presence of attractive nonlinearity; 
the opposite happens in the case of a repulsive interaction. This can also be appreciated from a Husimi representation of 
the modes (see fig.(\ref{hus-z})).

While for repulsive interactions the spreading of the distribution,
 together with peak damping, 
 seems to depend monotonically on the nonlinearity strength, 
the attractive case exhibits more complicated features (see fig.(\ref{hpm})).
Enhancement of the accelerator mode is only observed for 
small nonlinearities, while a striking feature appears 
at larger values of $|u|$, namely the accelerator 
mode is suppressed (see fig.(\ref{pac-x-ugr}a)). 
The intuitive explanation of this result 
is that strong focusing nonlinearity opposes to the 
separation of the wave packet into two parts;
indeed, in the case of exact resonance (namely $\tau=2\pi$), 
the mode is absent, so the whole wave freely falls without splitting and then 
the maximum height of the wave, plotted vs $u$ as in fig.(\ref{hpm}a), is 
found to monotonically increase to the left towards a saturation value.
While the behavior shown in fig.(\ref{hpm}) has been observed for a variety 
of other parameter choices, we mention that more complex, strongly fluctuating behaviour 
was sometimes observed at large focusing nonlinearities. In all such cases a bad correspondence 
between the quantum and the pseudoclassical dynamics was also observed, already in the linear case.

We remark that the mode damping is sensitive
to the choice of the initial state, as shown in fig.(\ref{pac-x-ugr}). 
While a gaussian initial wave packet leads to the mentioned QAM suppression, 
we may tailor a QAM enhancing initial condition as follows: we take the 
quasimomentum $\beta_0$ that in the linear case dominates the mode (here 
$\beta_0=\pi/\tau -\eta/2\simeq 0.5551$ \cite{FGR02}) 
and we drop from the initial 
gaussian all components with $|\beta -\beta_0 |> 0.15$. As quasimomentum 
is the fractional part of momentum, this leads to the comb like state 
of fig.(\ref{pac-x-ugr}b). Even through quasimomentum is not conserved 
due to nonlinearity, the QAM is strongly enhanced with respect to the linear 
case and the recoiling part 
is almost cancelled.

Another way of looking at the nonlinear evolution with techniques that are 
proper in the linear setting is to consider the distribution function 
over quasimomenta, defined by
\begin{equation}
\label{dist}
f(\beta,t)\,=\,\sum_{n=- \infty}^{+\infty}\, \left| \langle n+\beta | \psi(t) \rangle \right|^2.
\end{equation}
This distribution is stationary under linear evolution, 
its shape being determined by the choice of the initial state. 

\begin{figure}
  \includegraphics[width=9cm,angle=0]{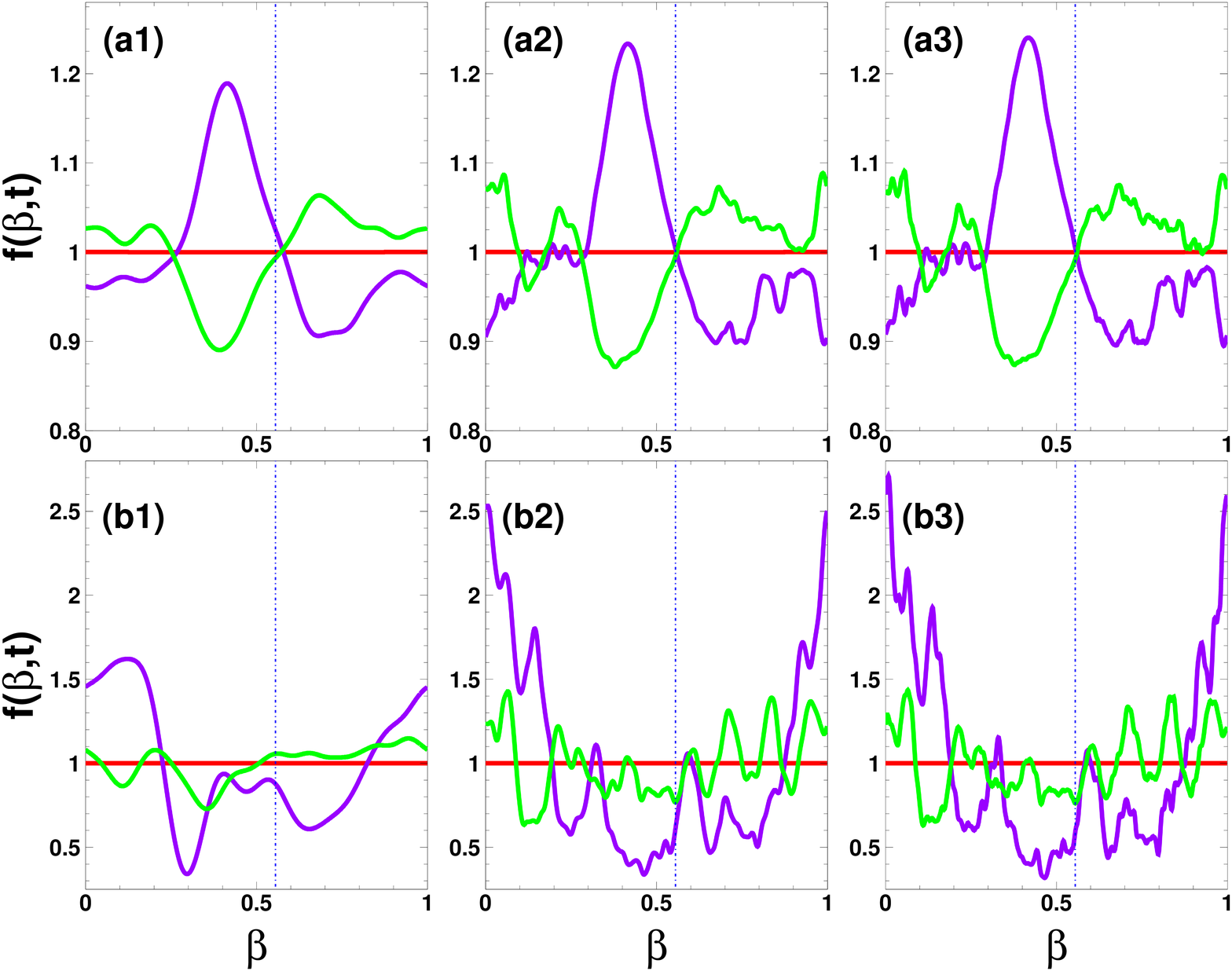}
  \caption{
(Color online) The quasi-momentum distribution function
(thick line) for $|u|=0.5$ (a) and 
$|u|=3$ (b) at
different times $(1) t=5$, $(2) t=25$ and $(3) t=45$.
Purple(dark)/green(light) lines refer to attractive/repulsive 
interactions.}\label{dqm-pr}
\end{figure}

We consider the evolution of a gaussian wave packet 
(for which the linear $f$ is essentially a constant - the 
horizontal red line of fig.(\ref{dqm-pr})), 
and probe the effect of nonlinearities of both signs. 
Typical results are as in fig.(\ref{dqm-pr}): the effect of attractive 
(repulsive) 
nonlinearity is to enhance (lower) the distribution around 
a value $\bar \beta\simeq 0.4$. 
No deviation occurs for quasimomentum $\beta_0$ (marked by vertical lines), 
whose wave function, according to fig.(\ref{pac-x-ugr}b), closely follows 
the linear pseudo-classical island. 
Again the $\bar \beta$ peak of the focusing case is suppressed 
for large focusing nonlinearities.

\begin{figure}
 \includegraphics[width=9cm,angle=0]{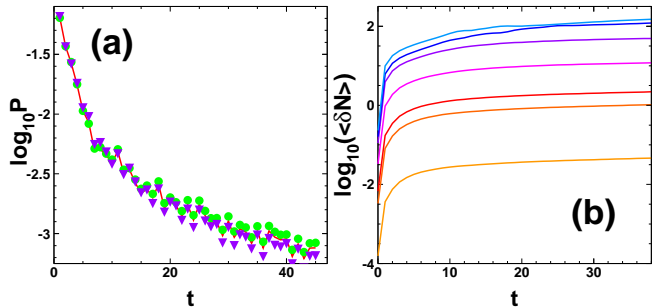}
 \caption{(Color online) (a) Probability inside the island for $|u|$=3 
(symbols and line as in fig.(\ref{pac1})). 
(b) The mean number of non-condensed particles 
vs the number of kicks, for $u$ equal to 
$0.1,0.5,0.75,1,2,5,7$ and
$10$ (starting from below); $12$ 
terms in the sum (\ref{ncon}) are considered.}\label{nconp}
\end{figure}

To make sure that our findings may be experimentally significant 
we discuss some stability issues: the first concerns 
decay properties of the QAMs. 
It is known that linear modes decay due to quantum tunnelling out 
of pseudoclassical islands \cite{sfgr}: 
we checked that, on the available time scale, the   
nonlinear decay behaves in a similar way. In fig.(\ref{nconp}a) the 
probability inside the classical 
island is shown as a function of time for the initial 
state of fig.(\ref{pac-x-ugr}b); it has been 
calculated integrating the Husimi distribution of each
 $\beta$-rotor fiber over the island area 
and summing the contributions of different rotors.

However in the condensate regime 
there is another possible mechanism that might completely modify
the former picture, namely depletion of the condensate due to proliferation of noncondensed, thermal particles. A standard technique to estimate the growth of the number of thermal particles is provided by the formalism of Castin and Dum \cite{CD}, which has been employed in similar contexts in \cite{CD-C}. To the lowest order in the perturbation expansion and in the limit 
of zero temperature $T\to 0$, the number of non-condensed particles 
is given by:
\begin{equation}
\label{ncon}
\langle \delta \hat N (t)\rangle =\sum_k \langle v_k (t) |v_k (t) \rangle
\end{equation}
where $v_k (t)$ is one of the mode functions of the system. 
The modal functions
 $(u_k(t),v_k (t))$ are pairs of functions that represent 
the time-dependent 
coefficients of 
the decomposition, in terms of annihilation and creation operators, of the 
equation of 
motion for the field operator describing the thermal excitations above 
the condensate. They describe the spatial dependence of these 
excitations and propagate by modified Bogoliubov equations.

Our findings (see fig.(\ref{nconp}b)) are consistent with a polynomial 
growth of noncondensed 
particles, namely in our parameter region 
(and within the time scale we typically consider) no exponential instability 
takes place. This is consistent with recent experimental work \cite{S}, 
where ${}^{87}$Rb atom condensate has been used to explore QAMs. 
In \cite{S}, a condensate of 50000 Rb atoms 
with repulsive interactions is realized. 
In the case of a "cigar shaped" trap, the relationship between the 
number of atoms 
in the condensate $N$ 
and the effective 1-d nonlinear coupling constant $u$ is, in our units, 
$N = u a_{\perp}^2/2 a_0$ \cite{confi}, where $a_0$ is the 
3-dimensional scattering length and 
 $a_\perp\gg a_0$ is the radial extension of the wave function. 
Using the parameter values of the experiment \cite{S}, one finds 
$N\simeq 10^5\cdot u$ and so $N\sim 50000$ corresponds to $u\sim 0.5$.  
Therefore our range of parameters includes the experimental accessible one.
 
 We have investigated effects of atomic interactions, 
in the form of a cubic nonlinearity, on the problem of quantum accelerator 
modes: in particular we have characterized the consequences of both 
attractive and repulsive interaction; we have also provided evidences that 
the modes are not strongly unstable when reasonable parameters are chosen.

We thank G. S. Summy for providing us with details of his work.
This work has been partially supported by the MIUR-PRIN 2005 
project "Transport properties of classical and quantum systems". 
S.F. acknowledges support by the Israel Science Foundation (ISF), by the US-
Israel Binational Science Foundation (BSF), by the Minerva Center of Nonlinear Physics
of Complex Systems, by the Shlomo Kaplansky academic chair. I.G. acknowledges hospitality 
by the Institute of Theoretical Physics at the Technion where part of this work was done.


\end{document}